\documentclass[sigconf,nonacm]{acmart}

\usepackage{algorithmic}
\usepackage{graphicx}
\usepackage{xcolor}
\usepackage{tabularx}
\usepackage{listings}
\usepackage{tcolorbox}
\usepackage{enumitem}
\usepackage{CJKutf8}
\usepackage{hyperref}

\lstdefinestyle{mystyle}{
    commentstyle=\color{codegreen},
    keywordstyle=\color{magenta},
    numberstyle=\tiny\color{codegray},
    stringstyle=\color{codepurple},
    basicstyle=\tiny\ttfamily,
    breakatwhitespace=false,
    breaklines=true,
    captionpos=b,
    keepspaces=true,
    numbers=left,
    numbersep=5pt,
    showspaces=false,
    showstringspaces=false,
    showtabs=false,
    tabsize=2,
    columns=fixed
}
\definecolor{codegray}{rgb}{0.5,0.5,0.5}
\definecolor{codegreen}{rgb}{0,0.6,0}
\definecolor{codepurple}{rgb}{0.58,0,0.82}

\newcommand{\result}[1]{%
\begin{tcolorbox}[colframe=black,boxrule=0.5pt,arc=4pt,
      left=6pt,right=6pt,top=6pt,bottom=6pt,boxsep=0pt,width=\columnwidth]%
      {\emph{#1}}
\end{tcolorbox}%
}

\AtBeginDocument{%
  }

\setcopyright{cc}
\copyrightyear{2026}
\acmConference[ASE'26 (NIER track)]{41st IEEE/ACM International Conference on Automated Software Engineering}{12--16 October 2026}{Munich, Germany}

\begin{document}

\title{Large Language Models as Empirical Computers}
\title{Empirical Computation: Prompting versus Programming}

\author{Eric Tang}
\authornote{Both authors contributed equally. Eric conducted the work as intern at MPI-SP.}
\affiliation{%
  \country{CMU, USA}
}

\author{Jing Liu}
\authornotemark[1]
\affiliation{%
  \country{MPI-SP, Germany}
}

\author{Marcel B{\"o}hme}
\affiliation{%
  \country{MPI-SP, Germany}
}

\renewcommand{\shortauthors}{Tang, Liu and B{\"o}hme}

\begin{abstract}
Large Language Models (LLMs) can solve \emph{any} computational problem \emph{without} an algorithm in a runtime \emph{independent} of the computational complexity of that problem. Instead of specifying precisely how to solve problem instance using \emph{programming}, we ask an LLM or agent to solve the problem instance using \emph{prompting}. Outputs are sampled from a distribution rather than generated procedurally.

In this vision paper, we explore the challenges and opportunities of this new form of computation and observe that its capabilities and limits \emph{cannot} be understood within the classic, rationalist framework of computation.
Hence, we appeal to the software engineering (SE) community to develop the foundations and techniques required to analyze the properties of this ``empirical computation'' as it generates solutions to computational problems:
How can we analyze and improve the correctness of LLMs solving a computational problem in the general, in the problem-specific, or in the instance-specific? What are the properties and fundamental limits of empirical computation? This paper aims to establish empirical computation as a field in SE that is timely and rich with interesting problems.

\end{abstract}








\maketitle

\section{Introduction}  
Large Language Models (LLMs) and LLM agents are poised to change the field of software engineering. Only two years ago Microsoft announced the first LLM-based code synthesis tool Co-Pilot \cite{copilot}. Now, \emph{25\% of all of Google's new code is LLM-generated} \cite{googlellm}---despite concerns about ``hallucination'' as a  source of bugs and security flaws \cite{insecureLLM,insecureLLM2}.

Today, we solve computational problems\footnote{In this paper, we take a very broad view of the term \emph{computational problem}. For instance, the computational problem that is solved by a given program is the relationship between the inputs and outputs of that program (or the ``purpose'' of that program).} by \emph{programming}: Using a programming language, we tell the machine precisely \emph{how} instances of that problem are to be solved. For instance, if our problem is to sort an array of numbers, we think about a \emph{specific} algorithm (e.g., merge sort) before developing a \emph{specific} program (or function) that implements it. The program will expect the input to be given in a \emph{specific} format, type, or data structure (e.g., \texttt{uint32\_t}). A program thus ``forces'' the problem instance to be provided in the specified format.
In software engineering, we have many approaches to analyze and even verify the correctness of such programs. We can localize and repair the underlying fault. Fundamentally, there exists no correct program that can solve the average instance of the sorting problem faster than $O(n\log n)$.

\newpage
In the future, we expect \emph{some} computational problems to be solved by \emph{prompting} (e.g., using an informal encoding as a natural language prompt). The developer tells the ``empirical computer'' (e.g., an LLM or an agentic system) informally about the instance of their computational problem, and the empirical computer returns the result---if needed in a machine-readable format. The result is returned as empirically most likely rather than verifiably correct. 

For our sorting problem, we can ask an LLM for instance to sort \begin{CJK*}{UTF8}{gbsn}\{一千一百二十三，九，负五，三点七\}\end{CJK*}\footnote{The Chinese characters specify the array \{1123, 9, -5, 3.7\}.} and to return, e.g., a JSON file with the sorted array. In our experiments we find that the time it takes the LLM to solve sorting problems increases linearly in array size ($O(n)$ instead of $O(n\log n)$). That runtime increases at all is likely due to the increasing context length. Yet, we find that the likelihood that the LLM returns the sorted list \emph{correctly} critically depends on the empirical likelihood of the problem instance:
\begin{itemize}[leftmargin=*]
\item While the LLM sorts arrays of length 50 correctly in over 90\% of cases, it sorts arrays of length 150 correctly only in 58\% of cases.
\item When \emph{reasoning} (or "thinking") capability is enabled, LLM can achieve near-perfect correctness (within the time or token limits).
\item While the LLM can sort the majority of arrays of length 20 correctly if the numbers are spelled out in English, it is far less reliable if numbers are spelled out in German.
\end{itemize}
We notice that, unlike for traditional programs, we cannot localize and repair the cause of an incorrectness once identified so that this and similar problem instances are solved correctly in the future.
In the experiments for this vision paper, we also report on results of four other computational problems of varying complexity.

\result{%
We argue that the capabilities and limits of empirical computation \emph{cannot} be understood within the classic, rationalist framework of computation and call on the software engineering community to develop new instruments for the analysis of empirical computation.
}\vspace{-0.1cm}

The purpose of this paper is to establish empirical computation as an emerging and interesting area in software engineering.
What are the properties of empirical computation? How to analyze the fundamental limits? Are there any guarantees? How can we analyze, estimate, or predict the correctness of empirical computation in the general, in the problem-specific, or in the instance-specific? 

\section{Motivation and Open Challenges}

\emph{Programming vs prompting to solve problems}.
Today, we write programs to solve problems. As mentioned, a program represents \emph{how} that problem is solved. Just like the problem can often be decomposed into smaller sub-problems, a program can be composed from many smaller computational units (e.g., functions). A program is a \emph{specific} solution for a \emph{specific} problem. The software engineering community has developed many approaches and techniques to analyze the properties (incl. correctness) of a program \cite{roadmap}.
However, in the future, we might simply prompt an agent to solve a problem. There is no program to analyze with respect to the problem it solves. \textbf{What can we say about the result? How do we \emph{test}~the~correct\-ness of agents for various computational problems}?

Programming gives us absolute, direct, and precise control over the correctness and efficiency of a computation. The procedure of the computation is explicit and predefined. The fault of an incorrect computation can be localized and repaired.
In contrast, prompting offers more flexibility, particularly when requirements are uncertain. Without constraints on language \cite{grammaraligned,chopchop}, input structure, or algorithm, an empirical computer returns the result after interpreting the user-provided, informal description of the problem statement using the available context. If we found the result to be incorrect, \textbf{how can we programmatically \emph{improve} correctness for future instances} (without introducing regressions)?\vspace{-0.1cm}

\paragraph{Formal vs informal representations of problem instances}
In \emph{programming}, inputs are provided using a \emph{specific} format, type, or data structure. Everyone who, as part of a software testing class, has been asked to implement and to test a program for the triangle classification problem knows that we have to be \emph{very precise} about that format. When parsing from a file, do we give lengths or angles, separated by commas or dashes? Are spaces and tabs allowed? When receiving as function input, should I use a primitive type, like (unsigned) integers,  or implement my own class? Are negative numbers allowed? What about floating point values?

In \emph{prompting}, the input can be provided informally in \emph{any} format. The correct interpretation arises from context, much like the meaning of a word may arise from the Wittgensteinian language game \cite{wittgenstein}. There is no predefined ``contract'' between caller and callee that specifies the precise format or data structure for the input. This offers flexibility at the interface, but at the cost of ambiguity.

We experimentally explore the correctness of empirical sorting when numbers are provided not using digits but expressed entirely differently, e.g., using words in German (zweiundvierzig [speak, two-and-forty; 42]) or as characters in Korean. We find that the result of empirical sorting is more likely correct if numbers are provided in a language that is well-represented on the internet.\vspace{-0.1cm}

\paragraph{Limits in terms of efficiency vs correctness}
The computational limits of programming to solve computational problems are well-studied in theoretical computer science. However, an LLM or agent finds answers to any computational problem in a time that aligns primarily with tokenization and runtime behavior rather than classical computational complexity. For instance, an LLM might take as much time to increment a number as it would to predict the plaintext from a given SHA-512 hash. Of course, correctness is a different matter. Are there problems that are particularly amenable to empirical computation?
\textbf{How can we quantify the limits of empirical computation in the general or in the problem-specific}?\footnote{We note that PAC learning \cite{pac} and learnability theory offers insight only on the requirements for (or dependence on) the training data (i.e., sample complexity).}

We experimentally study the efficiency and correctness of \emph{empirical computation} on simple, well-studied problems such as sorting and searching. In our measurements, runtime correlates primarily with tokenization and model/runtime behavior rather than classical problem complexity.  For example, in \emph{empirical sorting} or \emph{empirical searching} (sorted or unsorted lists), the baseline model returns the correct result for lists of 50 random numbers in $\approx 90\%$ of trials.\vspace{-0.1cm}

\paragraph{Formal vs informal notions of correctness}
In programming, it is upon the programmer to elicit and to understand the computational problem to be solved (i.e., the requirements) before implementing a program that is meant to solve \emph{all} instances of that problem. If the computational problem is precisely understood (such as sorting or searching), it can be encoded as formal specification, and given the specification, the programmer could formally verify, i.e., \emph{guarantee} the correctness of her program w.r.t. that specification.

However, sometimes the problem is not precisely understood and can only be described vaguely (cf. requirements elicitation). In such cases, prompting offers more flexibility and allows a contextual interpretation of a vaguely described computational problem (which may even be further elicited interactively). Yet, unlike in formal language, a natural language description can be ambiguous, and the way the prompt is designed can substantially impact the correctness of the response (cf. prompt engineering).
\textbf{What are effective ways to describe a computational problem to maximize correctness}?

\begin{figure*}
  \centering
  \includegraphics[width=0.98\textwidth]{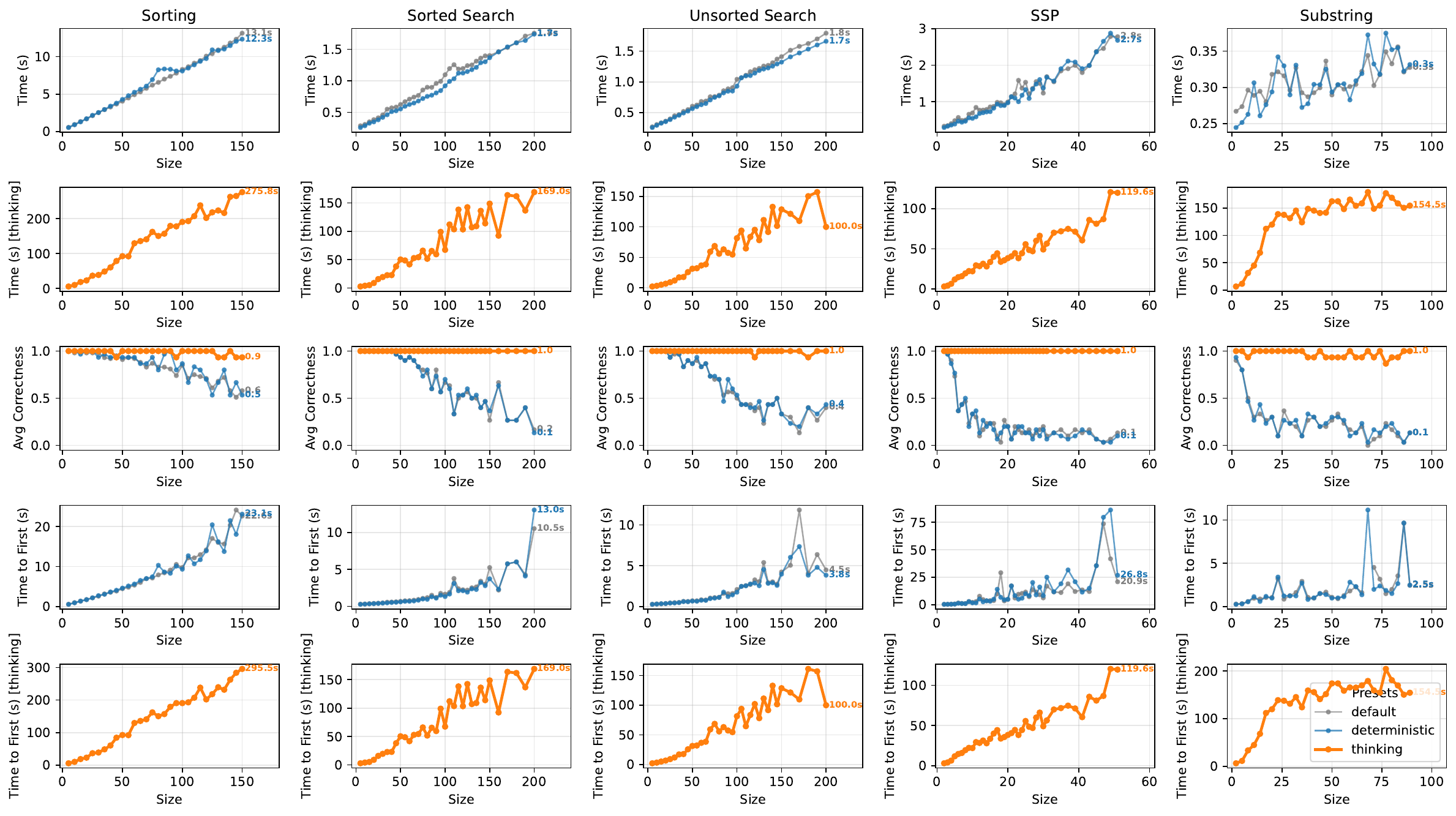}  
  \caption{LLM-performance on various computational problems across \emph{Default}, \emph{Deterministic} (both at least 30 repetitions), and \emph{Thinking} modes (15 repetitions). \emph{Top}: Average time to solution. \emph{Middle}: Average proportion of correct solutions. \emph{Bottom}: Expected time to generate the first correct solution.}
  \label{fig:overall_comparison}
\end{figure*}

\section{Preliminary Experiments}
To explore the challenges and opportunities of the empirical approach to computation, we conducted a large number of experiments. While we take a very broad view of the term "computational problem", here, for our experiments, we focus on simple, classic, well-studied problems, like sorting or searching, whose correctness is well-defined and computational complexity is well-studied.

\subsection{Experimental Setup}

\emph{Implementation}. For our evaluation, we developed 1.2k lines of Python code that prompts an LLM for various computational problems like sorting, for which the concrete \emph{prompt} is
\result{{``Sort the elements in the given collection in ascending order, and return only the sorted collection in the list format: $\langle$\texttt{numbers}$\rangle$.''}}
The input was provided as a Python list and formatted into a prompt string (\texttt{numbers}) according to problem-specific templates.
The output was parsed from the LLM response into (hopefully sorted) Python list.
In addition to sorting, we also conducted experiments on the following problems: \begin{itemize}[leftmargin=0.5cm]
\item searching sorted lists; $O(\log n)$,
\item searching unsorted lists; $O(n)$,
\item computing the longest palindromic substring; $O(n)$ \cite{manacher}, and
\item finding a subset with a given subset sum; $O(2^{\frac{n}{2}})$ \cite{horowitz}.
\end{itemize}
We chose to run our experiments with recent open-weights model \texttt{gemma-4-26b-a4b}~\cite{gemma4} locally to eliminate network latency and ensure accurate timing measurements. The model ranks higher than \texttt{DeepSeek-R1-0528}~\cite{deepseekr1} on the \emph{Artificial Analysis Intelligence Index}~\cite{ArtificialAnalysis2026} and is small enough to run on a MacBook Pro with Apple M5 Pro chip and 48GB unified memory. Based on the observed inference speed ($\sim$60 tokens/s), we set 600s timeout and 64k context length limit. We compare three configurations: \emph{Default} (temperature=1.0), \emph{Thinking} (reasoning mode enabled), and \emph{Deterministic} (temperature=0.0).

\paragraph{Methodology} For sorting, we generated Python lists of random numbers and of random length using existing functions from the standard Python library \texttt{random} (e.g., \texttt{random.sample}). As discussed in more detail in the individual sections, we varied properties like precision and magnitude of the numbers, as well as the length of the array. For every randomly generated input, we recorded the time taken (i.e., \emph{efficiency}) and whether the LLM-generated result was correct (i.e., \emph{correctness}).

For the other problems, we take a similar approach. For \emph{searching}, we generate a random or sorted list of numbers as above, select a random element to search for and prompt the LLM to return the corresponding index or negative one (-1). For the subset sum problem (\emph{SSP}), we generate a random list of numbers as above, find the sum $S$ of a random subset, and prompt the LLM to return that subset whose total is $S$. For the longest palindromic substring problem (\emph{substring}), we construct a random string of characters and prompt the LLM to return the result. For the latter two, we implement simple Python programs as ground truth. 

\subsection{Efficiency versus Correctness}

\emph{Efficiency}. \autoref{fig:overall_comparison}.top shows the efficiency of empirical computation, i.e., the average time it took the LLM to solve instances of various computational problems, as a function of instance size $n$. For four of five problems (except Substring) and all three modes, the time to find a solution is \emph{linearly} increasing in list size. The increase is likely due to the corresponding increase in the number of input and output tokens \cite{attention}, rather than the computational complexity of the problems. Note that the classic theory of computation predicts a difference between the average complexity of searching sorted versus unsorted lists. For Substring ($O(n)$), the execution time appears sublinear.

\emph{Correctness}.
\autoref{fig:overall_comparison}.middle shows the correctness of empirical computation, i.e., the proportion of empirically solved instances that are actually correct, as a function of $n$. Surprisingly, for \emph{thinking} mode correctness remains high (>0.9) and unaffected by the size of the input for the sizes, we could reasonably test.
For deterministic and default mode, we observe a clear decrease in correctness across problems. Interestingly, for sorting and both searching problems, the reduction in correctness in $n$ is roughly equivalent: For Python lists with 50 integers,
(a)~the probability that empirical sorting returns the correctly sorted list is equivalent to a coin flip (0.5), and
(b)~the probability that empirical searching returns the correct index is also equivalent to a coin flip (0.5).
In contrast, for the substring problem, correctness reduces to close to zero (0) already for strings with five characters likely due to tokenization. For the subset sum problem, 
correctness is close to zero (0) by lists of length 10. 
%
For these two modes, if we combine the efficiency and correctness results into the expected time until the LLM returns the first correct solution (Fig.~\ref{fig:overall_comparison}.bottom), we can clearly see the consequence of the decrease in correctness on the performance of empirical computation for these problems.

There might be several (non-traditional) mechanisms to improve the correctness of empirical computation, e.g., with methods like majority voting, 
prompt engineering~\cite{sahoo2024systematicsurveypromptengineering},
fine-tuning~\cite{han2024parameterefficientfinetuninglargemodels},
re\-inforce\-ment learning~\cite{kaelbling1996reinforcementlearningsurvey}, 
or test-time-training~\cite{Liang_2024}.
Currently, for our sorting problem, the LLM would sometimes add numbers not present in the input or remove numbers that were present. Some numbers would appear multiple times, and sometimes the list would not be sorted entirely. Large inputs (e.g., $n>200$) would sometimes be truncated or produce repeated sequences.

\begin{figure}
  \centering
  \includegraphics[width=0.83\columnwidth]{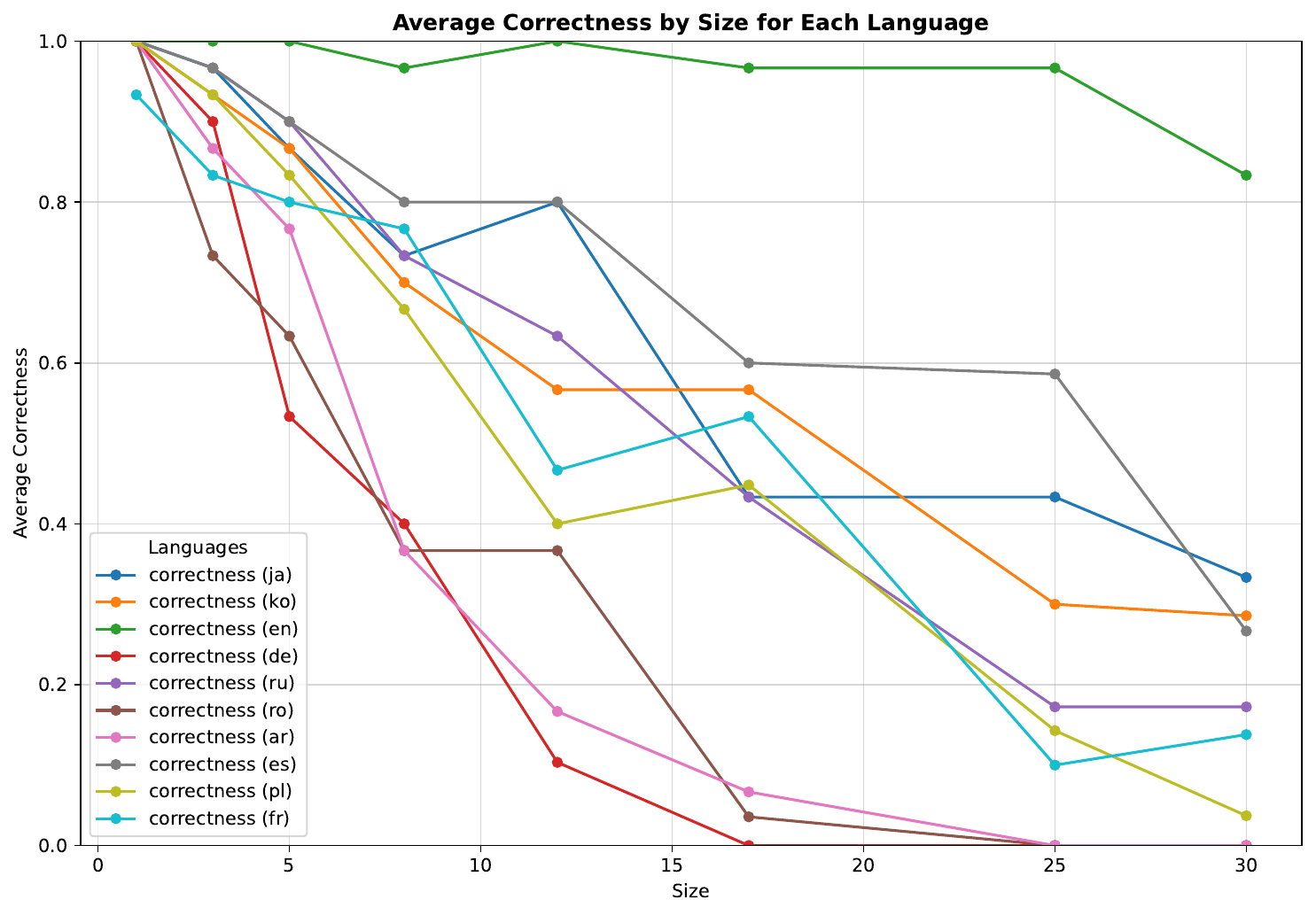}\vspace{-0.3cm}
  \caption{Correctness of empirical sorting across languages (i.e., the proportion of correctly sorted lists; default mode).\vspace{-0.2cm}}
  \label{fig:language}
\end{figure}

\subsection{Inputs in Natural Language}
A fundamental difference between the formal approach to computation and the empirical approach is that no ``formal contract'' is required for the latter on how the inputs are provided.
In contrast, a program requires inputs to be passed in a certain structure that is formally agreed on beforehand. That input structure could be a specific file format, like this PDF file, a specific data structure, like a linked list, graph, tree, or object, or it could be a primitive data type, like a (signed or unsigned) integer or float.

In contrast, an LLM takes inputs flexibly, either informally in natural language or formally, e.g., as structured JSON file. The LLM infers from context how the provided input is interpreted. This flexibility enables broader applicability but also introduces ambiguities. There is no certainty about how an input is interpreted.

\emph{Methodology}.
We used a Python library called \texttt{num2words} to translate the random numbers in the list into words or characters in different languages. The list of languages is shown on the top right in \autoref{fig:language}. We prompted the LLM (default mode) to sort the list of words and compared its output with the correct answer obtained by applying \texttt{num2words} to the sorted list of numbers.

\emph{Results}.
\autoref{fig:language} shows the correctness of empirical sorting, i.e., the proportion of empirically solved instances that are actually correct, as a function of $n$, when words or characters are used for numbers instead of digits. In comparison to \autoref{fig:overall_comparison}.middle, we can immediately see that the correctness of empirical sorting reduces more quickly. For languages that are well represented on the internet, correctness of empirical sorting is significantly better than for underrepresented languages. For instance, sorting a list of 20 numbers represented in English is correct with very high probability (>95\%), while when represented in German it is almost never correct (<1\%) in our 50 trials.

\section{Perspective and Vision}

\result{Everyone talks about trustworthy artificial intelligence. Yet, no one knows how to define, test, \textbf{improve}, or guarantee the correctness of the response of an LLM or an LLM agent to a given prompt. Our classic theory of computation offers no help in answering such questions; nor does the powerful program analysis machinery that we have developed in the software engineering community (e.g., static or dynamic analysis, verification, or model checking).\\
\hspace*{0.25cm} It is our \textbf{vision} that the analysis of the properties of empirical computation will emerge as a new area in software engineering that is both timely and rich with interesting problems.}

In the spirit of a research programme, we open several fundamental questions for the software engineering community.

\textbf{Empirical program analysis}. Given an empirical computer, like a system of LLM agents, what type of statements can we make about the correctness on various computational problems \cite{verigrey,clotho,AAAI26-incoherence} (whether or not formally specified \cite{specs1})? 
How can we extend existing formal methods, incl. verification, to work for empirical computers (beyond properties like robustness \cite{robustness})? In the absence of a program (i.e., a set of instructions) to analyze, we should develop tailored, scalable, blackbox empirical program analyses \cite{statistical,reachability,residualrisk,mutaflow} that proceed by statistical, counterfactual, or causal reasoning.

\textbf{Prompt Engineering}. What are the most effective encodings of a problem statement in natural language \cite{wei2023chainofthoughtpromptingelicitsreasoning}? Can we predict the correctness for a given problem instance \cite{AAAI26-incoherence}? How can we (reactively---or better proactively) improve correctness on specific computational problems or their instances in the absence of a (problem-specific) program? 

\textbf{Limits and capabilities of empirical computation}. How can we maximize the generality of our statements about the correctness of empirical computation? Can we make statements about correctness across \emph{all instances} of a problem? Can we make general statements about the correctness of an empirical computer, i.e., across all problems, whether known or unknown? What are the fundamental limits of empirical computation?

\textbf{Call to action}. The analysis of correctness and other properties is a classic problem in software engineering. Hence, we call on the software engineering research community to develop the  tools and techniques required to analyze the correctness of empirical computation as an important step toward the systematic analysis of the trustworthiness of artificial intelligence.

\section{Data Availability}

The implementation code, data, and scripts used in this study are available at \url{https://github.com/chinggg/EmpiricalComputation}.

\begin{acks}
This research was conducted as part of the CS@max planck internship program. This research is partially funded by the European Union (ERC project AT\_SCALE, 101179366). Views and opinions expressed
are however those of the author(s) only and do not necessarily reflect those of the European Union or the European Research Council Executive Agency. Neither the European Union nor the granting authority can be held responsible for them.
\end{acks}

\bibliographystyle{ACM-Reference-Format}
\bibliography{references}

\end{document}